\documentclass[12pt,a4paper]{article}
\usepackage[latin1]{inputenc}
\usepackage{amsmath}
\usepackage{placeins}
\usepackage{url}
\usepackage{natbib}
\usepackage{color,xcolor,setspace,geometry}
\setcitestyle{authoryear,open={(},close={)}}
\usepackage{cite}
\usepackage{amsfonts}
\usepackage{arydshln}
\usepackage{tikz}
\usetikzlibrary{shapes}
\usetikzlibrary{plotmarks}
\usetikzlibrary{tikzmark,matrix,calc}
\usetikzlibrary{arrows.meta,calc,decorations.markings,math,arrows.meta}
\usepackage{amssymb}
\usepackage{multirow}
\usepackage{subcaption}
\usepackage{graphicx}
\begin{document}
\begin{spacing}{1.35}

\title{Predicting play calls in the National Football League using hidden Markov models}

\author{
Marius \"Otting\thanks{Bielefeld University, Germany}
}

\date{}

\maketitle

\begin{abstract}
In recent years, data-driven approaches have become a popular tool in a variety of sports to gain an advantage by, e.g., analysing potential strategies of opponents. Whereas the availability of play-by-play or player tracking data in sports such as basketball and baseball has led to an increase of sports analytics studies, equivalent datasets for the National Football League (NFL) were not freely available for a long time. In this contribution, we consider a comprehensive play-by-play NFL dataset provided by \url{www.kaggle.com}, comprising 289,191 observations in total, to predict play calls in the NFL using hidden Markov models. The resulting out-of-sample prediction accuracy for the 2018 NFL season is 71.5\%, which is substantially higher compared to similar studies on play call predictions in the NFL.
\end{abstract}

\section{Introduction}
Unpredictability of play calls is widely accepted to be a key ingredient to success in the NFL. For example, according to several players of the 2017 Dallas Cowboys, being too predictable regarding their play calling may have been one reason for their elimination from the playoff contention of the 2017 NFL season. Being unpredictable hence is desirable, and, vice versa, it is clearly also of interest to be able to accurately predict the opponent's next play call. In earlier studies, play call predictions were carried out by simple arithmetics, such as calculating the relative frequencies of runs and passes of previous matches \citep{heiny2011predicting}. 
Driven by the availability of play-by-play NFL data, several studies considered statistical models for play call predictions. These studies can be divided in those where play-by-play data only is considered (see, \citealp{heiny2011predicting, teich2016nfl}) and those who consider additional data on the players on the field, such as the number of offensive players for a certain position and player ratings (see \citealp{leepredicting, joashpredicting}). The former report prediction accuracy of about 0.67, whereas the latter provide prediction accuracy of about 0.75.

However, most of these studies use basic statistical models, e.g.\ linear discriminant analysis, logistic regression, or decision trees, which do not account for the time series structure of the data at hand.
This contribution considers HMMs for modelling and forecasting NFL play calls. In the recent past, HMMs have been applied in different areas of research for forecasting, including stock markets (see, e.g., \citealp{de2013dynamic, dias2015clustering}), environmental science (see, e.g., \citealp{chambers2012earthquake, tseng2020forecasting}) and political conflicts \citep{schrodt2006forecasting}.
Within HMMs, the observations are assumed to be driven by an underlying state variable. In the context of play calling, the underlying states serve as a proxy for the team's current propensity to make a pass (as opposed to a run). The state sequence is modelled as a Markov chain, thereby inducing correlation in the observations and hence accounting for the time series structure of the data. HMMs are fitted to data from seasons 2009 to 2017 to predict the play calls for season 2018.
In practice, these predictions are helpful for defense coordinators to make adjustments in real time on the field. Offense coordinators may also benefit from these models, since they allow them to check the predictability of their own play calls. 

This paper is organised as follows: Section \ref{chap:data} describes the the play-by-play data and provides exploratory data analysis. Section \ref{chap:methods} explains HMMs in furhther detail, and section \ref{chap:results} presents the results.

\section{Data}\label{chap:data}
The data for predicting play calls in the NFL were taken from \url{www.kaggle.com}, covering (almost) all plays of regular season matches between 2009 to 2018. In total, $m = 2,526$ matches are considered\footnote{The data comprises 2,526 regular-season matches out of 2,560 matches which have taken place in the time period considered.}, each of which is split up into two time series (one for each team's offense), totalling in 5,052 time series containing 318,691 plays. The observed time series $\{y_{m,p}\}_{p=1,\ldots,P_m}$ indicates whether a run or a pass play has been called in the $p$-th play in match $m$, with

$$ 
y_{m,p} = 
\begin{cases}
1, & \text{if $p$--th play is a pass;} \\
0, & \text{otherwise}
\end{cases}
$$
and $P_m$ denoting the total number of plays in match $m$. For all matches considered, other plays such as field goals and kickoffs, which occur typically at the beginning or the end of drives, are ignored here. Since the main goal is to predict play calls, we divide the data into a training and a test data set. The data set for training the models covers all matches from seasons 2009 -- 2017, comprising 2,302 matches and 289,191 plays. The test data covers 224 matches, totalling in 29,500 plays. For the full data set, about 58.4\% of play calls were passes.

Since the play of the offense is likely affected by intermediate information on the match (such as the current score), several covariates are considered, which have also been considered by previous studies on predicting play calls summarised above: a dummy indicating whether the match is played at home (\textit{home}), the yards to go for a first down (\textit{ydstogo}), the current down number (\textit{down1}, \textit{down2}, \textit{down3}, and \textit{down4}), a dummy indicating whether the formation is shotgun (\textit{shotgun}), a dummy indicating whether the play is a no-huddle play (\textit{no-huddle}), the difference in the intermediate score (own score minus the opponent's score) (\textit{scorediff}), a dummy indicating whether the current play is a goal-to-go play (\textit{goaltogo}), and a dummy indicating whether the team is starting within 10 yards of their own end zone (\textit{yardline90}). Table \ref{tab:nfl_descriptives} summarises the covariates and displays corresponding descriptive statistics (for the full data set). 

\begin{table}[h] \centering 
  \caption{Descriptive statistics of the covariates.} 
  \label{tab:nfl_descriptives} 
  \scalebox{0.8}{
\begin{tabular}{@{\extracolsep{5pt}}lcccc} 
\\[-1.8ex]\hline 
\hline \\[-1.8ex] 
  & \multicolumn{1}{c}{mean} & \multicolumn{1}{c}{st.\ dev.} & \multicolumn{1}{c}{min.} & \multicolumn{1}{c}{max.} \\ 
\hline \\[-1.8ex] 
\textit{pass} (response) & 0.584 & 0.493 & 0 & 1 \\ 
\textit{home} & 0.503 & 0.500 & 0 & 1 \\ 
\textit{ydstogo} & 8.634 & 3.931 & 1 & 50 \\ 
\textit{down1} & 0.443 & 0.497 & 0 & 1 \\ 
\textit{down2} & 0.333 & 0.471 & 0 & 1 \\ 
\textit{down3} & 0.209 & 0.407 & 0 & 1 \\ 
\textit{down4} & 0.015 & 0.121 & 0 & 1 \\ 
\textit{shotgun} & 0.525 & 0.499 & 0 & 1 \\ 
\textit{no-huddle} & 0.087 & 0.282 & 0 & 1 \\ 
\textit{scorediff} & $-$1.458 & 10.84 & $-$59 & 59 \\ 
\textit{goaltogo} & 0.057 & 0.232 & 0 & 1 \\ 
\textit{yardline90} & 0.033 & 0.178 & 0 & 1 \\ 
\hline \\[-1.8ex] 
\end{tabular}}
\end{table} 

\begin{figure}
    \centering
    \includegraphics[scale = 0.75]{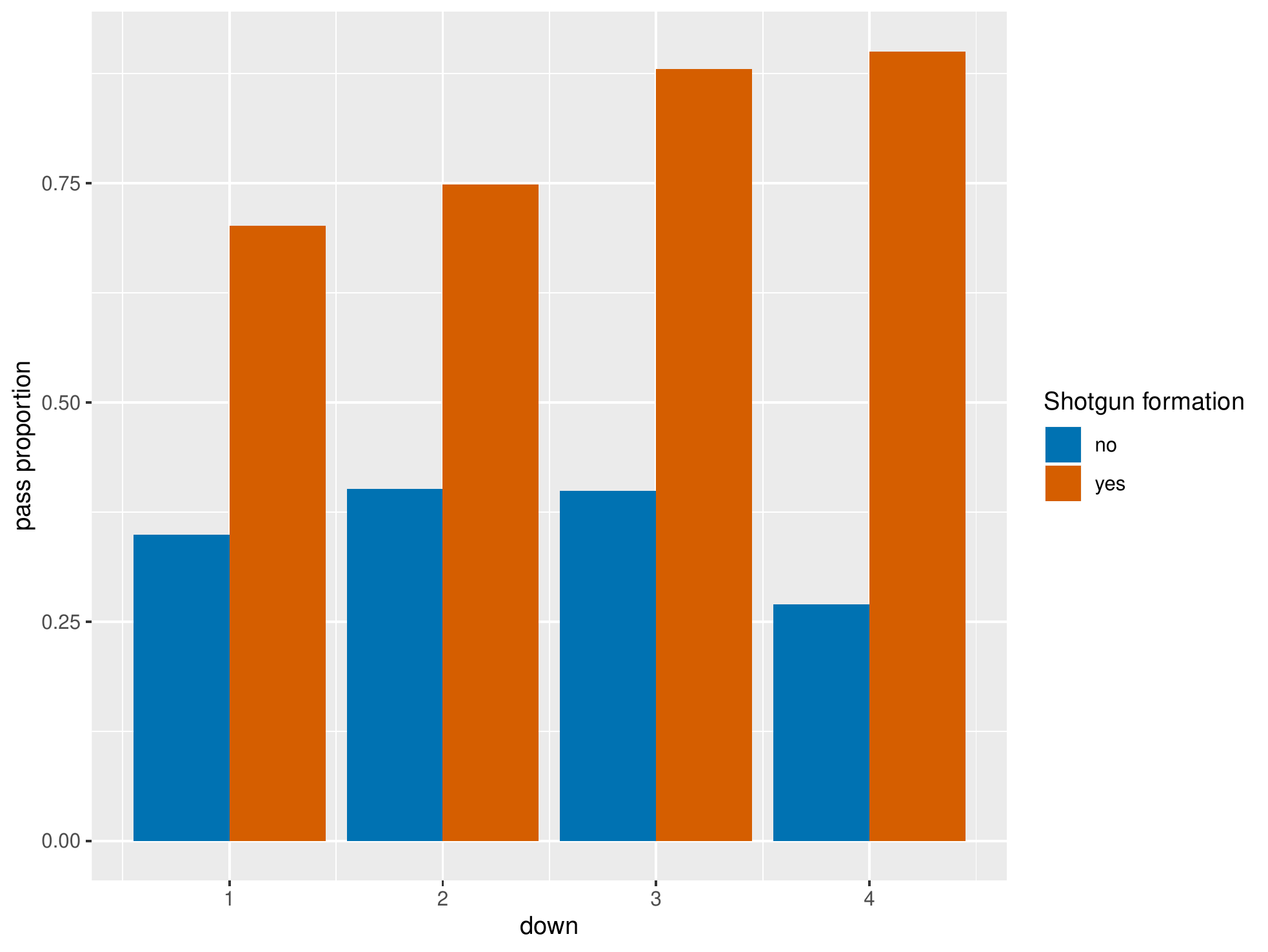}
    \caption{Empirical proportions for a pass found in the data for different downs and the shotgun formation.}
    \label{fig:down_shotgun}
\end{figure}

\begin{figure}
    \centering
    \includegraphics[scale = 0.75]{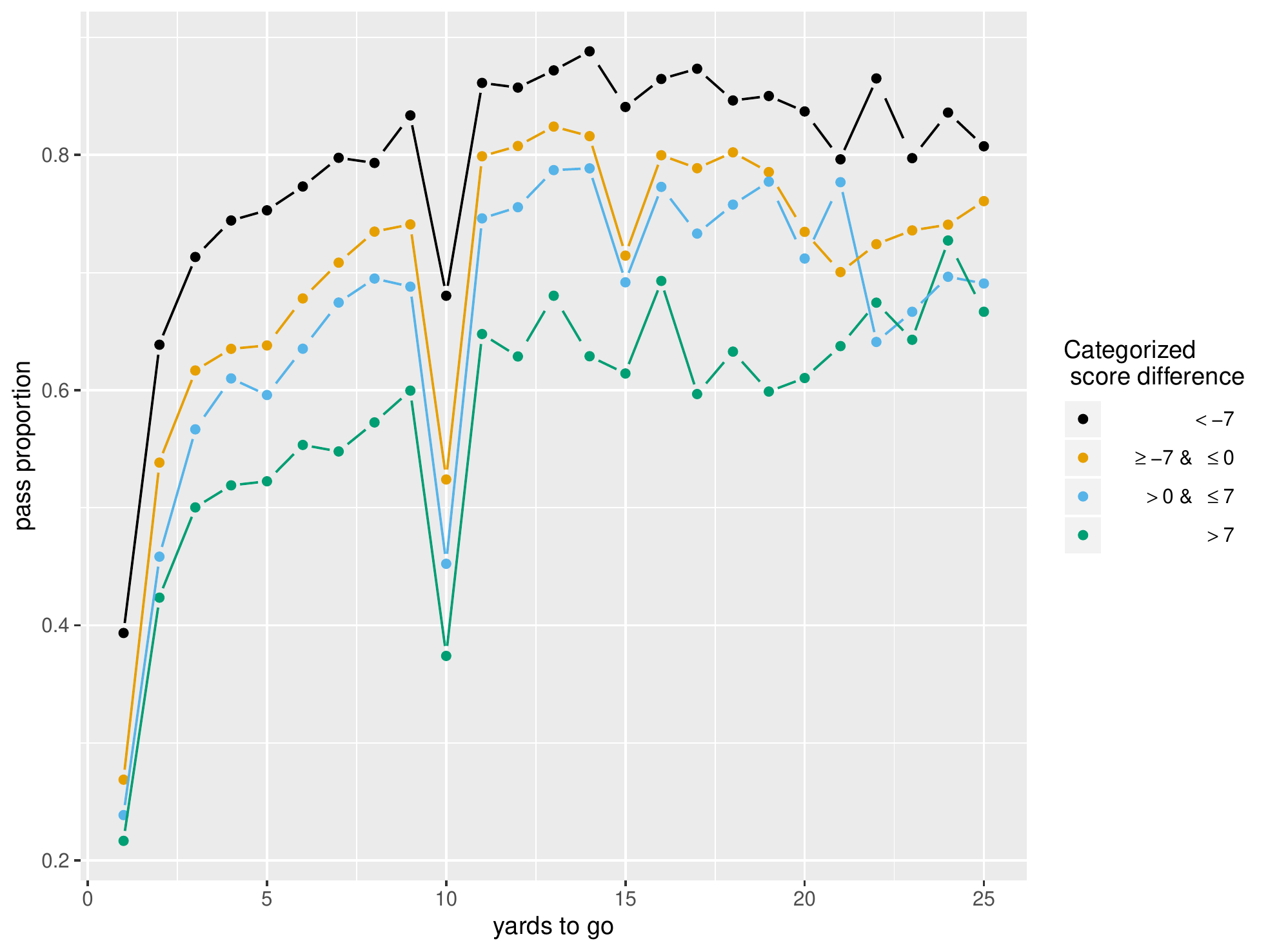}
    \caption{Empirical proportions for a pass found in the data for the different yards to go for a first down. Colours indicate the (categorised) score difference. The proportion for a pass for 10 yards to go is relatively low, since most of these observations correspond to a first down, where a run is more likely. Observations with more than 25 yards to go are excluded (the number of observations for each of these categories is less than 100).}
    \label{fig:scorediff}
\end{figure}

To investigate how the play calling varies with different downs and the shotgun formation, Figure \ref{fig:down_shotgun} shows the empirical proportions for a pass found in the data, separated for the different downs and the shotgun formation. As indicated by the figure, a pass becomes more likely with increasing number of downs, and there is a substantial increase in passes observed if the team is in shotgun formation. However, whether a run or a pass is called is also likely to depend on the yards to go for a first down, which is shown in Figure \ref{fig:scorediff}, indicating that a pass becomes more likely the more yards are needed for a first down. The colours in Figure \ref{fig:scorediff} indicate the (categorised) score difference, suggesting that a pass becomes more likely if teams are trailing. 

\begin{figure}
    \centering
    \includegraphics[scale = 0.9]{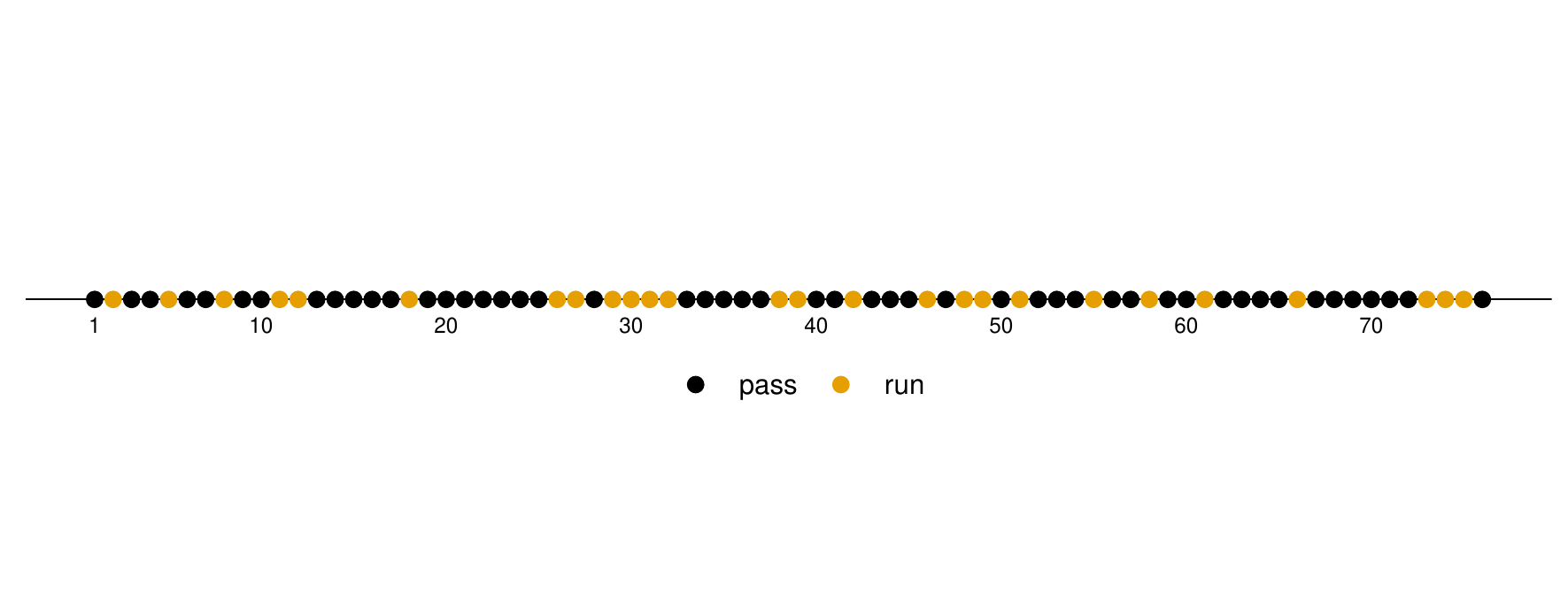}
    \caption{Example time series found in the data: the play calls of the New Orleans Saints observed for the match against the New York Giants played on November 1, 2015.}
    \label{fig:timeseries}\vspace*{-9pt}
\end{figure}

In addition to the covariates potentially affecting the decision to call a pass or a run, one example time series from the data set, corresponding to the play calls observed for the New Orleans Saints in the match against the New York Giants played in November 2015 is shown in Figure \ref{fig:timeseries}. With 101 points scored in total, this match is one of the highest scoring NFL games.
The plays shown in the figure underline that there are periods with a fairly high number of passing plays (e.g.\ around play 20), and those where more runs are called (e.g.\ around play 30).

\begin{table}[!htbp] \centering 
  \caption{Descriptive statistics of the covariates considered.} 
  \label{tab:descriptives} 
\begin{tabular}{@{\extracolsep{5pt}}lcccc} 
\\[-1.8ex]\hline 
\hline \\[-1.8ex] 
  & \multicolumn{1}{c}{mean} & \multicolumn{1}{c}{std.\ dev.} & \multicolumn{1}{c}{min} & \multicolumn{1}{c}{max} \\ 
\hline \\[-1.8ex] 
\textit{pass} (response) & 0.584 & 0.493 & 0 & 1 \\ 
\textit{home} & 0.503 & 0.500 & 0 & 1 \\ 
\textit{ydstogo} & 8.634 & 3.931 & 1 & 50 \\ 
\textit{down1} & 0.443 & 0.497 & 0 & 1 \\ 
\textit{down2} & 0.333 & 0.471 & 0 & 1 \\ 
\textit{down3} & 0.209 & 0.407 & 0 & 1 \\ 
\textit{down4} & 0.015 & 0.121 & 0 & 1 \\ 
\textit{shotgun} & 0.525 & 0.499 & 0 & 1 \\ 
\textit{no-huddle} & 0.087 & 0.282 & 0 & 1 \\ 
\textit{scorediff} & $-$1.458 & 10.84 & $-$59 & 59 \\ 
\textit{goaltogo} & 0.057 & 0.232 & 0 & 1 \\ 
\textit{yardline90} & 0.033 & 0.178 & 0 & 1 \\ 
\hline \\[-1.8ex] 
\end{tabular} 
\end{table}

\section{Modelling and forecasting play-calls}\label{chap:methods}
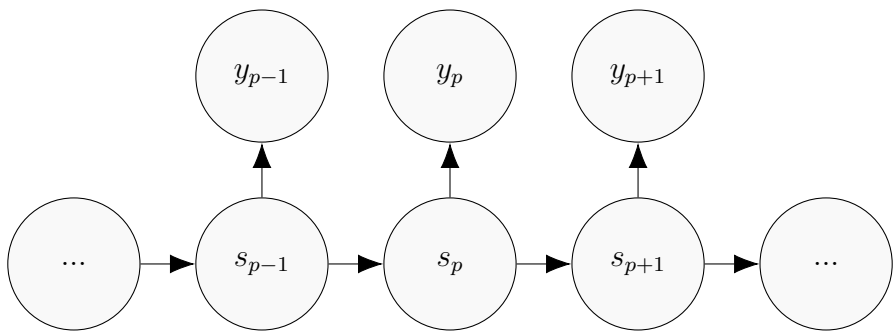
\begin{figure}[h!]
    \centering
	\begin{tikzpicture}
	\node[circle,draw=black, fill=gray!5, inner sep=0pt, minimum size=50pt] (A) at (2, -5) {$s_{p-1}$};
	\node[circle,draw=black, fill=gray!5, inner sep=0pt, minimum size=50pt] (A1) at (-0.5, -5) {...};
	\node[circle,draw=black, fill=gray!5, inner sep=0pt, minimum size=50pt] (B) at (4.5, -5) {$s_{p}$};
	\node[circle,draw=black, fill=gray!5, inner sep=0pt, minimum size=50pt] (C) at (7, -5) {$s_{p+1}$};
	\node[circle,draw=black, fill=gray!5, inner sep=0pt, minimum size=50pt] (C1) at (9.5, -5) {...};
	\node[circle,draw=black, fill=gray!5, inner sep=0pt, minimum size=50pt] (Y1) at (2, -2.5) {$y_{p-1}$};
	\node[circle,draw=black, fill=gray!5, inner sep=0pt, minimum size=50pt] (Y2) at (4.5, -2.5) {$y_{p}$};
	\node[circle,draw=black, fill=gray!5, inner sep=0pt, minimum size=50pt] (Y3) at (7, -2.5) {$y_{p+1}$};
	\draw[-{Latex[scale=2]}] (A)--(B);
	\draw[-{Latex[scale=2]}] (B)--(C);
	\draw[-{Latex[scale=2]}] (A1)--(A);
	\draw[-{Latex[scale=2]}] (C)--(C1);
	\draw[-{Latex[scale=2]}] (A)--(Y1);
	\draw[-{Latex[scale=2]}] (B)--(Y2);
	\draw[-{Latex[scale=2]}] (C)--(Y3);
	\end{tikzpicture}
\caption{Dependence structure of the HMM considered. Each observation $y_{p}$ is 
assumed to be generated by one of $N$ distributions according to the state process 
$s_{p}$, which serves for the team's current propensity to make a pass (as opposed to a run).}
\label{fig:HMM}
\end{figure}

To account for the periods of passes and runs as indicated by Figure \ref{fig:timeseries}, HMMs are considered for modelling and forecasting play calls. The underlying states can be interpreted as the propensity to make a pass (as opposed to a run) of the team considered. 
A HMM involves two components, namely an observed state-dependent process and an unobserved Markov chain with $N$ states, assuming that the observations are generated by one of $N$ pre-specified state-dependent distributions. The dependence structure of the HMM considered is shown in Figure \ref{fig:HMM}. Here, the observed time series are the play calls $\{y_{m,p}\}_{p=1,\ldots,P_m}$, which are denoted from now on by $y_p$ for notational simplicity. The unobserved state process, modelled by a $N$-state Markov chain, is denoted by $\{s_p\}_{p=1,\ldots,P_m }$. For the state transitions, a transition probability matrix (t.p.m.) $\boldsymbol{\Gamma} = (\gamma_{ij})$ is defined, with $\gamma_{ij}=\Pr(s_p = j | s_{p-1}=i$), i.e.\ the probability of switching from state $i$ at play $p-1$ to state $j$ in play $p$. 
For the model formulation of an HMM to be completed, the number of states $N$ and the class of the state-dependent distribution have to be selected. Since the play calls are binary, the Bernoulli distribution is chosen here. The corresponding probabilities of the observation given state $i$, i.e.\ $f(y_p\, |\, s_p = i)$ are comprised in the $i-$th diagonal element of the $N \times N$ diagonal matrix $\mathbf{P}(y_{p})$. Since assuming a team to start in its stationary distribution at the beginning of an American football match is fairly unrealistic, we estimate the initial distribution $\boldsymbol{\delta}= \big(\Pr (s_{p} = 1),\ldots,\Pr (s_{p} = N) \big)$.

To include the covariates introduced above which may lead to state-switching, we allow the transition probabilities $\gamma_{ij}$ to depend on covariates at play $p$. This is done by linking $\gamma_{ij}^{(p)}$ to covariates (denoted by $x_1^{(p)},\ldots,x_k^{(p)}$) using the multinomial logit link:
$$
\gamma_{ij}^{(p)} = \dfrac{\exp(\eta_{ij}^{(p)})}{\sum_{k=1}^N \exp(\eta_{ik}^{(p)})}
$$
with \vspace{0.5cm}
$$ \eta_{ij}^{(p)} = 
\begin{cases}
\beta_0^{(ij)} + \sum_{l=1}^K \beta_l^{(ij)} x_l^{(p)}  & \text{if }\, i\ne j; \\
0 & \text{otherwise}.
\end{cases} 
\vspace{1cm}
$$
Since the transition probabilities depend on covariates, the t.p.m.\ as introduced above is not constant across time, and hence denoted by $\boldsymbol{\Gamma}^{(p)}$.
To formulate the likelihood, we apply the forward algorithm, which allows to calculate the likelihood recursively at low computational cost \citep{zucchini2016hidden}. The likelihood for a single match $m$ is then given by: 

\begin{equation*}
L = \boldsymbol{\delta} \mathbf{P}(y_{m,1}) \boldsymbol{\Gamma}^{(m,2)}\mathbf{P}(y_{m,2}) \dots \boldsymbol{\Gamma}^{({m,P_m})}\mathbf{P}(y_{m,P_m}) \mathbf{1}
\end{equation*}
with column vector $\mathbf{1}=(1,\ldots,1)' \in \mathbb{R}^N$ \citep{zucchini2016hidden}. 
To obtain the likelihood for the full data set, we assume independence between the individual matches such that the likelihood is given by the product of likelihoods for the individual matches:

\begin{equation*}
L = \prod_{m=1}^{M} \boldsymbol{\delta} \mathbf{P}(y_{m,1}) \boldsymbol{\Gamma}^{(m,2)}\mathbf{P}(y_{m,2}) \dots \boldsymbol{\Gamma}^{({m,P_m})}\mathbf{P}(y_{m,P_m}) \mathbf{1},
\end{equation*}
where $M$ denotes the total number of matches.
The model parameters are estimated by numerically maximising the likelihood using \texttt{nlm()} in R \citep{rcoreteam}. Subsequently, we predict play calls for the test data using the fitted models. Specifically, to forecast play calls, the forecast distribution is considered, which is for a single match given as a ratio of likelihoods (dropping the subscript $m$ for notational simplicity):

$$
\Pr(y_{P+1} = y \,|\, \mathbf{y}^{(P)}) = \dfrac{\boldsymbol{\delta} \mathbf{P}(y_{1}) \boldsymbol{\Gamma}^{({2})} \mathbf{P}(y_{2}) \cdots \boldsymbol{\Gamma}^{({P})} \mathbf{P}(y_{P}) \boldsymbol{\Gamma}^{(y)} \mathbf{P}(y) \mathbf{1}}{\boldsymbol{\delta} \mathbf{P}(y_{1}) \boldsymbol{\Gamma}^{({2})} \mathbf{P}(y_{2}) \cdots \boldsymbol{\Gamma}^{({P})} \mathbf{P}(y_{P}) \mathbf{1}},
$$
where $\boldsymbol{\Gamma}^{(y)}$ and $\mathbf{y}^{(P)}$ denote the t.p.m.\ as implied by the new covariates and the vector of all preceding observations of the match considered, respectively \citep{zucchini2016hidden}. The play which is most likely under the forecast distribution is then taken as the one-step-ahead forecast. 
To address heterogeneity between teams, the models are fitted to data of each team individually instead of pooling the data of all teams. The corresponding results are presented in the next section.

\section{Results}\label{chap:results}
Before presenting the results on the prediction of play calls, the number of states $N$ and the covariates have to be selected. As the number of parameters (due to the inclusion of covariates) increases considerably fast compared to the number of observations per team, we select $N=2$ states here to avoid numerical instability. We apply a forward selection of the covariates described in Section \ref{chap:data} based on the AIC. In addition, we also include several interactions between the covariates, such as an interaction between \textit{ydstogo} and \textit{scorediff}, which was already indicated  by in Figure \ref{fig:scorediff}. Based on further explanatory data analysis, the following additional interaction terms are considered: interactions between the different downs and \textit{ydstogo}, between \textit{shotgun} and \textit{ydstogo}, between \textit{nohudlle} and \textit{scorediff}, and between \textit{nohuddle} and \textit{shotgun}. The AIC-based forward covariate selection is then applied for each team individually, with the covariates selected being slightly different between the teams.

The play call forecasts are evaluated by the prediction accuracy (i.e.\ the proportion of correct predictions), the precision (i.e.\ the proportion of predicted runs/passes that were actually correct) and the recall (i.e.\ the proportion of actual runs/passes that were identified correctly). The weighted average of the prediction accuracy over all teams is obtained as 0.715. This is a substantial improvement compared to existing studies that were also based on play-by-play data only (i.e.\ without including information on the players on the field). Moreover, the prediction accuracy obtained here is only slightly lower than the ones reported by \citet{leepredicting} and \citet{joashpredicting} (which are about 75\%), notably \textit{without} taking into account information about the players on the field.

\begin{figure}[!t]
    \centering
    \includegraphics[width=0.99\textwidth]{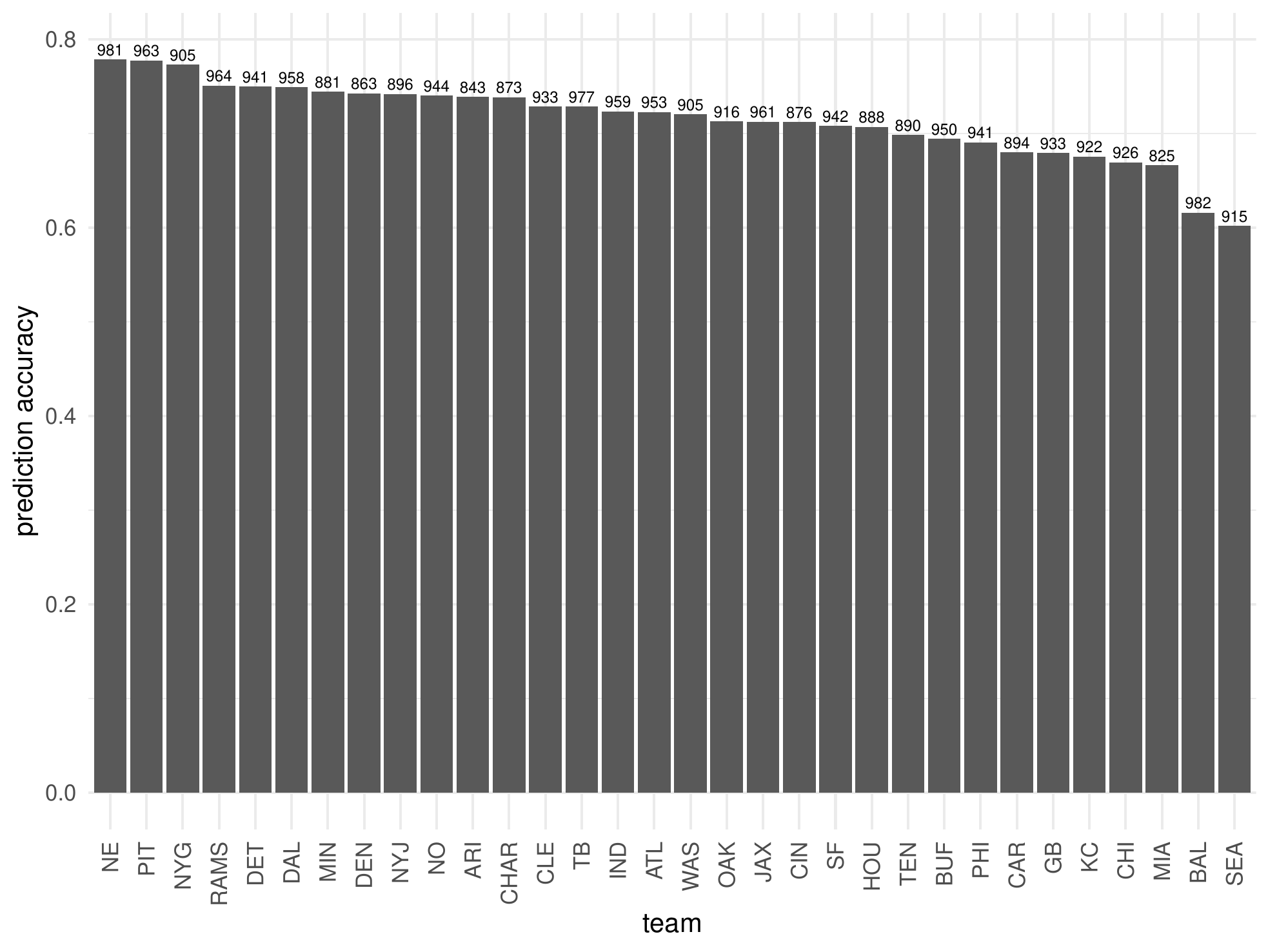}
    \caption{Prediction accuracy for the individual teams. The number of out-of-sample observations (i.e.\ of predicted plays) is shown at the top of the bars.}
    \label{fig:predteams}\vspace*{-9pt}
\end{figure}

The prediction accuracy for the individual teams is shown in Figure \ref{fig:predteams}, indicating that the lowest and highest prediction accuracy are obtained for the Seattle Seahaws (0.602) and the New England Patriots (0.779), respectively. In addition, the precision rates for a run range from 0.532 (Green Bay Packers) to 0.763 (Houston Texans), which can be interpreted as follows:\ when our model predicts a run for the Houston Texans (Green Bay Packers), it is correct in about 76.3\% (53.2\%) of all predicted runs. The recall rates for a run range from 0.324 (Baltimore Ravens) to 0.886 (Los Angeles Rams) --- in other words, our model correctly predicts 88.6\% of all runs for the Los Angeles Rams. For passing plays, precision and recall range from 0.559 (Seattle Seahawks) to 0.9 (Los Angeles Rams), and from 0.664 (Los Angeles Rams) to 0.922 (Pittsburgh Steelers), respectively. These summary statistics on the predicted play calls reveal that there are substantial differences in the predictive power with regard to the individual teams. Section \ref{chap:discussion} discusses practical implications following from these summary statistics. 
It took us on avarage 7 hours to conduct the AIC-based forward selection for the covariates on a standard desktop computer. However, using the fitted models to predict play calls takes less than a second for a single match, thus rendering the approach considered suitable for application in practice.

\section{Discussion}\label{chap:discussion}

The use of HMMs to predict play calls in the NFL indicates that the accuracy of the predictions is increased -- compared to similar previous studies -- by accounting for the time series structure of the data. 
We split the data into a training set (seasons 2009--2017) and a test set (season 2018), and fitted HMMs to the (training) data of all teams individually, which yields 71.5\% correctly predicted out-of-sample play calls. The prediction accuracy for the individual teams range from 60.2\% to 77.9\%, with the highest prediction accuracy obtained for the New England Patriots (see Figure \ref{fig:predteams}). 

Practitioners have to take into account the variation in the prediction accuracy across teams and plays. For example, if a pass is predicted for the Los Angeles Rams, it is fairly likely that the actual play will indeed be a pass (according to our model), since the corresponding precision is obtained as 90\%. On the other hand, if a pass is predicted for the  Seattle Seahawks, this forecast has to be treated with caution, as the precision is obtained as 55.9\%.
Additional aspects for practitioners are the costs of an incorrect decision. For example, if teams want to avoid that a pass is anticipated although the actual play of the opponent's offense is a run, then coaches should carefully consider the corresponding precision rates. Since the models presented here provide probabilistic forecasts and not only binary classifications, coaches could consult the forecasts only if the predicted probability exceeds a chosen threshold.
In any case, practitioners should not regard these models as a tool which delivers defense adjustments for each play automatically, but rather as an additional help to make better defense and offense plays, respectively.

Further research could focus on including additional covariates to improve the predictive power, such as the personnel of the team, i.e.\ the information on how many running backs/fullbacks, tight ends and wide receiver are on the field. In addition, the current strength of the team is not captured yet. This could be quantified by, for instance, the player ratings provided by the video game Madden, which was also done by \citet{leepredicting} and \citet{joashpredicting}. However, it is at least questionable whether information on players can indeed be used on the field in practice, since players are substituted fairly frequently during a match. Finally, updating the model throughout the 2018 season dynamically, rather than using the model fitted up to season 2018 in the out-of-sample prediction would further improve the predictive power.

\newpage

\bibliographystyle{apalike}
\bibliography{refs}

\end{spacing}
\end{document}